\documentstyle[12pt,epsfig]{article}
\def\titlefoot#1{\stepcounter{footnote}\footnotetext{#1}}
\newcommand{\Aname}[2]{ #1$^{\ref{#2}}$}

\begin{document}
\begin{center}
{\large EUROPEAN ORGANIZATION FOR NUCLEAR RESEARCH}
\end{center}
\begin{flushright}
CERN-EP/98-047\\
23 March 1998
\end{flushright}

\vspace{1.5cm}
\begin{center}
{\large \bf SEARCH FOR A NEW GAUGE BOSON IN $\pi^{0}$ DECAYS}
\end{center}
\begin{center}
{\large The NOMAD Collaboration}
\end{center}

\vspace{1.5cm}

\begin{abstract}

A search was made for a  new light gauge  boson $X$ which might be produced  in 
$\pi^{0}\rightarrow\gamma + X$ decay from neutral pions generated by 450-GeV
protons in the CERN SPS neutrino target. The X's would  penetrate the 
downstream shielding and be observed in the NOMAD detector 
via the Primakoff effect, in the  process of  $X \rightarrow\pi^{0}$ conversion
in the external Coulomb field of a nucleus.\
With  $1.45\times10^{18}$ protons on target, 20 candidate events with 
energy between 8 and 140 GeV  were found from the analysis of neutrino data.\
  This number is in agreement with the expectation  of  18.1$\pm$2.8 background events from standard neutrino processes.\ 
A new 90 $\% ~C.L.$ upper limit on the branching ratio $Br(\pi^{0}\rightarrow\gamma + X)<$ (3.3~to~1.9)$\times10^{-5}$ for $X$ masses ranging from 0 to 120 MeV/c$^{2}$ is obtained.\
\end{abstract}
{\small

\vspace{2.cm}
\begin{center}
{ \it Submitted to Physics Letters B}
\end{center}
\newpage
\begin{center}
{\large The NOMAD Collaboration}
\end{center} 
{
%
\Aname {J.~Altegoer}{Dortmund}
\Aname {P.~Astier}{Paris}
\Aname {D.~Autiero}{CERN}
\Aname {A.~Baldisseri}{Saclay}
\Aname {M.~Baldo-Ceolin}{Padova}
\Aname {G.~Ballocchi}{CERN}
\Aname {M.~Banner}{Paris}
\Aname {S.~Basa}{Lausanne}
\Aname {G.~Bassompierre}{LAPP}
\Aname {K.~Benslama}{Lausanne}
\Aname {N.~Besson}{Saclay}
\Aname {I.~Bird}{Lausanne}$^{,\ref{CERN}}$
\Aname {B.~Blumenfeld}{Johns Hopkins}
\Aname {F.~Bobisut}{Padova}
\Aname {J.~Bouchez}{Saclay}
\Aname {S.~Boyd}{Sydney}
\Aname {A.~Bueno}{Harvard}
\Aname {S.~Bunyatov}{Dubna}
\Aname {L.~Camilleri}{CERN}
\Aname {A.~Cardini}{UCLA}
\Aname {A.~Castera}{Paris}
\Aname {P.W.~Cattaneo}{Pavia}
\Aname {V.~Cavasinni}{Pisa}
\Aname {A.~Cervera-Villanueva}{CERN}
\Aname {G.M.~Collazuol}{Padova}
\Aname {G.~Conforto}{Urbino}
\Aname {C.~Conta}{Pavia}
\Aname {M.~Contalbrigo}{Padova}
\Aname {R.~Cousins}{UCLA}
\Aname {D.~Daniels}{Harvard}
\Aname {A.~De~Santo}{Pisa}
\Aname {T.~Del~Prete}{Pisa}
\Aname {T.~Dignan}{Harvard}
\Aname {L.~Di~Lella}{CERN}
\Aname {E.~do Couto e Silva}{CERN}
\Aname {I.J.~Donnelly}{Sydney}$^{,\ref{ANSTO}}$
\Aname {J.~Dumarchez}{Paris}
\Aname {T.~Fazio}{LAPP}
\Aname {G.J.~Feldman}{Harvard}
\Aname {R.~Ferrari}{Pavia}
\Aname {D.~Ferr\`ere}{CERN}
\Aname {V.~Flaminio}{Pisa}
\Aname {M.~Fraternali}{Pavia}
\Aname {J-M.~Gaillard}{LAPP}
\Aname {P.~Galumian}{Lausanne}
\Aname {E.~Gangler}{Paris}
\Aname {A.~Geiser}{CERN}
\Aname {D.~Geppert}{Dortmund}
\Aname {D.~Gibin}{Padova}
\Aname {S.N.~Gninenko}{INR}
\Aname {J-J.~Gomez-Cadenas}{Amherst}$^{,\ref{CERN}}$
\Aname {J.~Gosset}{Saclay}
\Aname {C.~G\"o\ss ling}{Dortmund}
\Aname {M.~Gouan\`ere}{LAPP}
\Aname {A.~Grant}{CERN}
\Aname {G.~Graziani}{Florence}
\Aname {A.~Guglielmi}{Padova}
\Aname {C.~Hagner}{Saclay}
\Aname {J.~Hernando}{Amherst}$^{,\ref{UCSC}}$
\Aname {D.~Hubbard}{Harvard}
\Aname {P.~Hurst}{Harvard}
\Aname {N.~Hyett}{Melbourne}
\Aname {E.~Iacopini}{Florence}
\Aname {C.~Joseph}{Lausanne}
\Aname {D.~Kekez}{Zagreb}
\Aname {B.~Khomenko}{CERN}
\Aname {M.M.~Kirsanov}{INR}
\Aname {O.~Klimov}{Dubna}
\Aname {J.~Kokkonen}{CERN}
\Aname {A.V.~Kovzelev}{INR}
\Aname {V.~Kuznetsov}{Dubna}
\Aname {S.~Lacaprara}{Padova}
\Aname {A.~Lanza}{Pavia}
\Aname {L.~La Rotonda}{Calabria}
\Aname {M.~Laveder}{Padova}
\Aname {A.~Letessier-Selvon}{Paris}
\Aname {J-M.~Levy}{Paris}
\Aname {L.~Linssen}{CERN}
\Aname {A.~Ljubi\v{c}i\'{c}}{Zagreb}
\Aname {J.~Long}{Johns Hopkins}$^{,\ref{Colorado}}$
\Aname {A.~Lupi}{Florence}
\Aname {E.~Manola-Poggioli}{LAPP}$^{,\ref{CERN}}$
\Aname {A.~Marchionni}{Florence}
\Aname {F.~Martelli}{Urbino}
\Aname {X.~M\'echain}{Saclay}
\Aname {J-P.~Mendiburu}{LAPP}
\Aname {J-P.~Meyer}{Saclay}
\Aname {M.~Mezzetto}{Padova}
\Aname {S.R.~Mishra}{Harvard}
\Aname {G.F.~Moorhead}{Melbourne}
\Aname {L.~Mossuz}{LAPP}
\Aname {P.~N\'ed\'elec}{LAPP}$^{,\ref{CERN}}$
\Aname {Yu.~Nefedov}{Dubna}
\Aname {C.~Nguyen-Mau}{Lausanne}
\Aname {D.~Orestano}{Rome}
\Aname {F.~Pastore}{Rome}
\Aname {L.S.~Peak}{Sydney}
\Aname {E.~Pennacchio}{Urbino}
\Aname {J-P.~Perroud}{Lausanne}
\Aname {H.~Pessard}{LAPP}
\Aname {R.~Petti}{Pavia}
\Aname {A.~Placci}{CERN}
\Aname {H.~Plothow-Besch}{CERN}
\Aname {A.~Pluquet}{Saclay}
\Aname {G.~Polesello}{Pavia}
\Aname {D.~Pollmann}{Dortmund}
\Aname {B.G.~Pope}{CERN}
\Aname {B.~Popov}{Dubna}$^{,\ref{Paris}}$
\Aname {C.~Poulsen}{Melbourne}
\Aname {P.~Rathouit}{Saclay}
\Aname {C.~Roda}{CERN}$^{,\ref{Pisa}}$
\Aname {A.~Rubbia}{CERN}
\Aname {F.~Salvatore}{Pavia}
\Aname {D.~Scannicchio}{Pavia}
\Aname {K.~Schahmaneche}{Paris}
\Aname {B.~Schmidt}{Dortmund}
\Aname {A.~Sconza}{Padova}
\Aname {M.~Serrano}{Paris}
\Aname {M.E.~Sevior}{Melbourne}
\Aname {D.~Sillou}{LAPP}
\Aname {F.J.P.~Soler}{Sydney}
\Aname {G.~Sozzi}{Lausanne}
\Aname {D.~Steele}{Johns Hopkins}$^{,\ref{Lausanne}}$
\Aname {P.~Steffen}{CERN}
\Aname {M.~Steininger}{Lausanne}
\Aname {U.~Stiegler}{CERN}
\Aname {M.~Stip\v{c}evi\'{c} }{Zagreb}
\Aname {T.~Stolarczyk}{Saclay}
\Aname {M.~Tareb-Reyes}{LAPP}$^{,\ref{Lausanne}}$
\Aname {G.N.~Taylor}{Melbourne}
\Aname {S.~Tereshchenko}{Dubna}
\Aname {A.N.~Toropin}{INR}
\Aname {A-M.~Touchard}{Paris}
\Aname {S.N.~Tovey}{Melbourne}
\Aname {M-T.~Tran}{Lausanne}
\Aname {E.~Tsesmelis}{CERN}
\Aname {J.~Ulrichs}{Sydney}
\Aname {V.~Uros}{Paris}
\Aname {M.~Valdata-Nappi}{Calabria}$^{,\ref{Perugia}}$
\Aname {V.~Valuev}{Dubna}$^{,\ref{LAPP}}$
\Aname {F.~Vannucci}{Paris}
\Aname {K.E.~Varvell}{Sydney}$^{,\ref{ANSTO}}$
\Aname {M.~Veltri}{Urbino}
\Aname {V.~Vercesi}{Pavia}
\Aname {D.~Verkindt}{LAPP}
\Aname {J-M.~Vieira}{Lausanne}
\Aname {T.~Vinogradova}{UCLA}
\Aname {M-K.~Vo}{Saclay}
\Aname {S.A.~Volkov}{INR}
\Aname {F.~Weber}{Harvard}
\Aname {T.~Weisse}{Dortmund}
\Aname {M.~Werlen}{Lausanne}
\Aname {F.~Wilson}{CERN}
\Aname {L.J.~Winton}{Melbourne}
\Aname {B.D.~Yabsley}{Sydney}
\Aname {H.~Zaccone}{Saclay}
\Aname {K.~Zuber}{Dortmund}
}
{\footnotesize
\titlefoot{Univ. of Massachusetts, Amherst, MA, USA \label{Amherst}}
\titlefoot{LAPP, Annecy, France \label{LAPP}} 
\titlefoot{Johns Hopkins Univ., Baltimore, MD, USA \label{Johns Hopkins}}
\titlefoot{Harvard Univ., Cambridge, MA, USA \label{Harvard}}
\titlefoot{Univ. of Calabria and INFN, Cosenza, Italy \label{Calabria}}
\titlefoot{Dortmund Univ., Dortmund, Germany \label{Dortmund}}
\titlefoot{JINR, Dubna, Russia \label{Dubna}}
\titlefoot{Univ. of Florence and INFN,  Florence, Italy \label{Florence}}
\titlefoot{CERN, Geneva, Switzerland \label{CERN}}
\titlefoot{University of Lausanne, Lausanne, Switzerland \label{Lausanne}}
\titlefoot{UCLA, Los Angeles, CA, USA \label{UCLA}} 
\titlefoot{University of Melbourne, Melbourne, Australia \label{Melbourne}}
\titlefoot{Inst. Nucl. Research, INR Moscow, Russia \label{INR}}
\titlefoot{Univ. of Padova and INFN, Padova, Italy \label{Padova}}
\titlefoot{LPNHE, Univ. of Paris, Paris VI and VII, France \label{Paris}}
\titlefoot{Univ. of Pavia and INFN, Pavia, Italy \label{Pavia}}
\titlefoot{Univ. of Pisa and INFN, Pisa, Italy \label{Pisa}}
\titlefoot{Roma-III Univ., Rome, Italy \label{Rome}}
\titlefoot{DAPNIA, CEA Saclay, France \label{Saclay}}
\titlefoot{Univ. of Sydney, Sydney, Australia \label{Sydney}}
\titlefoot{ANSTO,Sydney, Menai, Australia \label{ANSTO}}
\titlefoot{Univ. of Urbino, Urbino, and INFN Florence, Italy \label{Urbino}}
\titlefoot{Rudjer Bo\v{s}kovi\'{c} Institute, Zagreb, Croatia \label{Zagreb}}
\titlefoot{Now at Perugia Univ., Perugia, Italy \label{Perugia}}
\titlefoot{Now at UCSC, Santa Cruz, CA, USA \label{UCSC}}
\titlefoot{Now at Univ. of Colorado, Boulder, CO, USA \label{Colorado}}
}
\vfill 

\section{Introduction}

Many extensions of the Standard Model such as  GUTs~\cite{1}, 
super-symmetric~\cite{2}, super-string models~\cite{3} and models including a
new long-range interaction, i.e. the fifth force~\cite{4}, predict an extra
U$^{'}$(1) factor and therefore the existence of a new gauge boson $X$
corresponding to this new group.\ 
The predictions for the  mass of the  $X$ boson are not very firm and it could
be light enough ($M_{X}\ll M_{Z}$) for searches  at low energies.\ 

If the mass $M_{X}$ is of the order of the pion mass, an effective search could
be conducted for this new vector boson in the radiative decays of neutral 
pseudoscalar mesons $P\rightarrow\gamma + X$, where $P = \pi^{0},\eta$, or 
$\eta^{\prime}$, because  the decay rate of  $P\rightarrow\gamma~+~$ 
$\it any~new~particles~with~spin~0~or~\frac{1}{2}$ proves to be negligibly 
small~\cite{5}.\
Therefore, a positive result in the direct search for these decay modes could be
interpreted unambiguously as the discovery of a new light spin 1 particle, in
contrast with other experiments searching for light weakly interacting particles
in rare K, $\pi$ or $\mu$ decays~\cite{5,6}. 


From the analysis of the data from earlier experiments, constraints on the
branching ratio for the decay of  $P\rightarrow\gamma + X$ range from 
$10^{-7}$ to $10^{-3}$ depending on whether $X$ interacts with both quarks and
leptons or only with quarks.\ Since in the first case $X$ is a short lived
particle decaying mainly to $e^{+}e^{-}$ or $\nu \overline{\nu}$ pairs,
we will only consider the second case, where $X$ is a relatively long-lived
particle~\cite{6}.   

Direct searches for a signal from $\pi^{0}\rightarrow\gamma + X$ decay have been
performed in a few experiments with two different methods: i) searching for a
peak in inclusive photon spectra from two-body 
$\pi^{0}\rightarrow\gamma+nothing$ decays, where ``nothing'' means that $X$ is
not detected because it either has a long life time or decays into
$\nu\overline{\nu}$ pairs ( ref.~\cite{7,8,9}); and 
ii) searching  for a peak in the invariant mass spectrum of $e^{+}e^{-}$ pairs
from $\pi^{0}$ decays, which corresponds to the decay 
$X\rightarrow e^{+}e^{-}$~\cite{10}.\

The best experimental limit on the decay $\pi^{0}\rightarrow\gamma + X$ was
obtained recently by the Crystal Barrel Collaboration at CERN \cite{9}.
Using $\it p\overline{p}$ annihilations as a source of neutral pions,
they searched for a single peak in the inclusive photon energy spectrum.\
The branching ratio limit of (6~to~3)$\times 10^{-5}$  ($90\% C.L.$) was 
obtained for  $65<M_{X}<125~MeV$.\ A less stringent upper limit 
$BR(\pi^{0}\rightarrow\gamma+X) <$ (3~to~0.6)$\times 10^{-4}$ was
obtained for the mass region $0<M_{X}<65~MeV/c^{2}$.\ This result is 
valid  for the case where $X$ is a long-lived particle or it decays mainly to
$\nu \overline{\nu}$ pairs.\

In this paper, we present a more sensitive  upper limit on the branching ratio 
of the decay $\pi^{0}\rightarrow\gamma + X$ obtained by using a new 
method~\cite{11} from the analysis of high energy neutrino data taken by 
the NOMAD experiment at CERN.

\section{The NOMAD Detector}

The NOMAD detector, designed to search for a neutrino oscillation signal in 
the CERN SPS wide-band neutrino beam, is described in detail in ref.\cite{12}.\
A sketch of the NOMAD detector is shown in Figure~\ref{fig:1}.\  It consists of
a number of subdetectors most of which are located inside a 0.4 $T$ dipole 
magnet  with a  volume of 7.5$\times$3.5$\times$3.5 m$^{3}$.\  The relevant 
features for the present study will be briefly mentioned.\ 

The complete active target consists of 44 drift chambers (DC) mounted in 11
modules~\cite{13}.\  The target is followed by a transition radiation 
detector~(TRD) to enhance  $e/\pi$ separation and by a lead-glass 
electromagnetic calorimeter~(ECAL) with an upstream preshower detector~(PRS).
Five additional drift chambers are installed in the TRD region.

 Each of the 9 TRD modules~\cite{14} consists of  a radiator followed by a 
detection plane of vertical straw tubes.

The ECAL consists of 875  lead-glass Cerenkov counters of TF1-000 type arranged
in a matrix of 35 rows by 25 columns.\ Each counter is about 19 radiation 
lengths deep.
The energy resolution  is $\sigma/E = 0.01 + 0.032/\sqrt{E}$, where E is the 
shower energy in GeV.\ A more detailed description of the ECAL is given 
elsewhere~\cite{15}.

 The PRS is composed of two planes of proportional tubes (286 horizontal and
288 vertical tubes) preceded by a 9 mm thick (1.6 X$_{0}$) lead plate, providing
a spatial resolution $\sigma_{x,y} \approx 1 cm/\sqrt{E}$, where E is the shower
energy in GeV. The fiducial mass of this detector is about 700 kg.

The trigger for neutrino interactions in the target is provided by two planes of
scintillation counters T$_{1}$ and T$_{2}$. A veto  in front of the magnet 
rejects upstream neutrino interactions and muons incident on the detector.\
Neutrino interactions in the PRS or ECAL are collected by a  
$\overline{T_{1} \times T_{2}} \times ECAL$
trigger specially designed for this purpose. The ECAL signal is obtained as 
the OR of all counter signals exceeding a threshold of about 0.8 GeV.
 The timing of this signal  depends on the deposited energy. For energies above 
3 GeV a time resolution of a few ns and a trigger efficiency of  100$\%$ are
obtained.\ The average rate of the $\overline{T_{1} \times T_{2}} \times ECAL$
trigger is about 3/ 10$^{13}$ protons on the neutrino target (p.o.t.).\ 

 A hadronic calorimeter(HCAL) and a set of 10 drift chambers  placed behind
the magnet provide an estimate of the energy of the hadronic component in
the event and  muon identification, respectively.\ The HCAL is an 
iron-scintillator sampling calorimeter  consisting of 11 iron plates,
4.9 cm thick, separated by 1.8 cm gaps in which scintillator paddles 3.6 m long 
1 cm thick, and 18.3 cm wide are installed. The HCAL active area is 3.6 m wide
by 3.5 m high, and approximately 3.1 hadronic interaction lengths deep.

\section{Method of Search}

If the  decay $\pi^{0}\rightarrow\gamma+X$ exists, one expects a  flux of 
high energy $X$ bosons from the SPS neutrino target, since $\pi^{0}$'s are
abundantly produced in the forward direction  by 450 GeV protons either 
in the beryllium  target or in the beam dump following the decay tunnel.\
If $X$ is a long-lived particle, this flux  would penetrate the downstream 
shielding without significant attenuation  and would be observed in the NOMAD 
detector via the Primakoff effect, namely in the conversion process 
$X\rightarrow \pi^{0}$ in the external Coulomb field of a 
nucleus~\cite{11} (see Figure 2).\ 

Because the cross section for $X \rightarrow \pi^{0}$ conversion is proportional
to Z$^{2}$, we searched for these events in the lead of the preshower detector.\
The experimental signature of $X \rightarrow \pi^{0}$ conversion is a single 
high energy $\pi^{0}$ decaying into two photons  which results in a single 
isolated electromagnetic shower in the ECAL. The corresponding ECAL cluster 
should be matched to the PRS cluster from the converted photons and must not be
accompanied by a significant activity in any of the other NOMAD subdetectors.\
The occurrence of $X\rightarrow  \pi^{0}$ conversion  would appear as an excess
of neutrino-like interactions in the PRS with pure electromagnetic final 
states above those expected from  Monte Carlo predictions  of standard neutrino 
interactions. 

The expected energy spectrum of $X$-bosons at the NOMAD detector is shown in 
Figure 3 for a mass  $M_{X}$=10 MeV.\
The energy spectra of $\pi^{0}$'s produced in the neutrino target and in the 
beam dump have been obtained  with the same detailed GEANT \cite{17} simulation
used to predict the neutrino flux distributions at the NOMAD detector.\ 
The $X$-spectrum is considerably harder than that of neutrinos which  
is also shown for comparison.

\section{ Data Sample and Event Selection}

This analysis is based on the data taken during the first half of the neutrino
run in 1995, in which the NOMAD target was only partially installed and 
consisted of four drift chamber modules placed upstream of the TRD detector.\
The integrated number of protons delivered to the neutrino target during this
period was about 1.45 $\times 10^{18}$.\

Monte Carlo simulations use the LEPTO 6.1 event generator \cite{18}  supplemented by 
event generators for resonance, quasi-elastic and coherent neutrino processes 
and a full detector 
simulation based on GEANT\cite{17}.\ These simulations together with direct 
measurements of subdetector occupancy during the neutrino spill with a random
trigger \cite{12} show that $\nu$  and  $X \rightarrow \pi^{0}$ events  in the
PRS are accompanied by no significant activity in the DC and TRD.\
For this reason a simple cut on the number of hits $N_{hit}^{DC}$, 
$N_{hit}^{DC(TRD)}$ and $N_{hit}^{TRD}$ in the DC, in the DC located in the TRD
region, and in the TRD, respectively, were used as a veto to suppress events
from neutrino interactions which occurred in these regions or in the magnet
coils.
However, neutrino events, mainly with a neutral final state, which occurred in
the near upstream PRS region can also  pass these veto cuts and can be taken as
neutrino events in the PRS.\
For this reason the upstream region was also included in the simulations and
surviving events were considered as neutrino events in the PRS.\ 

The selection criteria for  $X\rightarrow \pi^{0}$ events are based on a full
Monte Carlo simulation of $X \rightarrow\pi^{0}$ conversions in the NOMAD 
detector and rely on their properties as mentioned in Section 3, namely: 
\begin{itemize}
\item
the quality of the match between the ECAL shower and the corresponding PRS cluster;
\item
the total energy and the shower shape  in the ECAL;
\item
the amount of energy deposition in the HCAL.
\end{itemize} 
Candidate events  were identified by the following  simultaneous
requirements:\\ 
   
\begin{itemize}

\item 
   $\overline{DC} \times \overline{TRD}$ (no activity in the DC or TRD ): 
$N_{hit}^{DC} \le 10$, $N_{hit}^{DC(TRD)} \le 4$, $N_{hit}^{TRD} \le 2$.\\

\item $PRS \times ECAL (> 8~ GeV)$: 
isolated PRS cluster matched to isolated ECAL cluster in both  X and Y planes.

At most two ECAL clusters in the ECAL were allowed. The energy of the most 
energetic cluster had to be  $E_{ECAL} > 8~ GeV$ and the second cluster 
had to be less than 0.3 GeV in order to reject pile-up events and ECAL noise.
The shape of the most energetic cluster was fitted to the shape expected from 
an electromagnetic shower and the  $\chi^{2}$ of the fit was required to be less
than 20~\cite{19}.
The differences $\Delta X$ and  $\Delta Y$ between the $X$ and $Y$ coordinates 
of the cluster in the PRS and the corresponding cluster in the ECAL were 
required to be
$|\Delta X| < 2~cm$,  $|\Delta Y| < 1.5~cm$.\

These conditions were used to identify isolated electromagnetic showers in the 
ECAL that originated from the conversion of photons from $\pi^{0}$'s 
produced in the preshower.\\

\item 

$\overline{HCAL}$ (no HCAL activity): the total visible energy was required to
be less than the HCAL noise threshold $E_{HCAL} < 0.4~ GeV$.\

This cut serves  to identify electromagnetic energy in the ECAL.\
The reliability of the HCAL veto to suppress conventional neutrino events with
an energy leakage  to the HCAL detector was checked with straight through
muons and with a selected sample of  $\nu_{\mu}$ charged current (CC) events
in the PRS.\\ 

\item
$\overline{MUON}$ : no track(s) in the muon chambers matched to the PRS cluster.

\end{itemize}

After applying these cuts to the initial sample of $4.83\times10^5$  events 
recorded with the ECAL trigger we found 20 candidate events for
$X \rightarrow \pi^{0}$ conversion.
The amount of background in this sample from standard  $\nu_{\mu}$ and
$\nu_{e}$ interactions was evaluated using the Monte Carlo (see Section 5).\  


Applying the
$\overline{DC}\times \overline{TRD}\times PRS\times ECAL (> 1 GeV)\times MUON$
selection criteria  to the same initial sample we find 3691 events
from $\nu_{\mu}CC$ interactions in the PRS. Here, $MUON$ denotes
single muon tracks extrapolated back from the muon chambers to the PRS and 
matched to the PRS cluster. The spectra  of energy deposited in the ECAL for 
these selected  $\nu_{\mu}CC$ events and for simulated events which pass the
same reconstruction program and selection cuts are found to be in agreement.
The $\overline{T_{1} \times T_{2}} \times ECAL$ trigger efficiency was obtained
from a Monte Carlo simulation and was found to be 71$\%$ and 97$\%$ for 
$\nu_{\mu}CC$ and $X \rightarrow \pi^{0}$ events, respectively.
Taking the overall selection efficiency into account, the total number of
$\nu_{\mu}CC$ events interacting in the PRS was found to be 
$N_{\nu_{\mu}CC}$=2.29$\times10^{4}$.\ 
The number of $\nu_{\mu} CC$ events was converted to the number of protons
on target $N_{pot}$ using the simulation of the neutrino flux at the NOMAD
detector and the known cross section value for $\nu_{\mu} CC$ interactions, 
giving
$N_{pot} = 1.36\times 10^{18}$.\ This number agrees within 10\% with the value
of $N_{pot}$ measured by  the beam monitors and was used further for 
normalisation.   The main uncertainty results from the contribution of events
from neutrino interactions in the downstream  TRD region which pass the DC and
TRD cuts and from backscattering in events occurring in the PRS.
By varying the  $\overline{DC}$ and $\overline{TRD}$ cuts, it was found that
the systematic error in the number of $\nu_{\mu}CC$ events in the PRS resulting
from backscattering is of the order of 10$\%$.

\section{ Background Events}

The main background to $X\rightarrow \pi^{0}$ conversions 
is expected from the following neutrino  processes occurring in the PRS or
upstream PRS region with a significant electromagnetic component in
the final state and with no significant energy deposition in the HCAL :

\begin{itemize}
\item $\nu_{\mu}CC$ interactions 
classified as muonless because the muon was not detected;

\item inclusive $\pi^{0}$ production from $\nu_{\mu}$ neutral current (NC) interactions; 

\item coherent  and  diffractive $\pi^{0}$ production;

\item quasi-elastic $\nu_{e}$ scattering.  

\item $\nu_{e}CC$ interactions;
\end{itemize}
  
Backgrounds from the trident events, 
$\nu_{e} + Pb\rightarrow \nu_{e} + e^{+} + e^{-} + Pb$  
and from pile-up events were found to be negligible.

To evaluate the amount of background in the data sample, simulated events were
processed through the same reconstruction program and selection criteria that
were used for real neutrino data. Then, all background distributions were 
summed up, taking into account the corresponding normalisation factors. These 
factors were calculated from beam composition and cross sections of the 
well-known processes listed above.
The total number of events interacting in the PRS and the number of expected 
candidate events after applying the selection criteria 
are given in Table 1 for each background process.
  
\begin{table}[hbt]
\begin{center}
\caption{Overall background estimate}
\begin{tabular}{|c|c|c|c|c|c|}
\hline
   item         & $\nu_{\mu}CC$ & $\nu_{\mu}NC$ & Coh.$\pi^{0}$ & $\nu_{e}-QE$ & $\nu_{e}CC$ \\
\hline
Total number of &               &               &               &              &             \\
interactions    &  22886        &      6866     &   59          &  6.2         & 320         \\
in the PRS      &               &               &               &              &             \\
\hline 
Number of       &               &               &               &              &             \\
expected        &   1.7$\pm$1.0 &  6.6$\pm$1.3  & 1.5$\pm$0.4   & 2.1$\pm$0.2  & 6.2$\pm$1.2 \\
candidate events&               &               &               &              &             \\
\hline 
\end{tabular}
\end{center}
\end{table}
The total background in the data sample was estimated to be 18.1$\pm$2.8 events,
where  statistical and systematic errors are added in quadrature. The fraction
of neutrino interactions in the PRS which satisfy all cuts is
6 $\times$ 10$^{-4}$.

\section{ Result}

Figure 4 shows the combined background and candidate event energy spectra 
in the ECAL.\  The agreement between data and Monte Carlo is good.\
The overall efficiency for $X \rightarrow \pi^{0}$ conversion detection was
found to be $\approx$24$\%$.\
The inefficiency  is mostly due to the requirement that at least one photon from
$\pi^{0}$ decay convert in the PRS.\

By subtracting the number of expected background events from the number of 
candidate events we obtain $N_{X\rightarrow \pi^{0}}$ = 1.9$\pm$5.3 events
showing no excess of $X \rightarrow \pi^{0}$ conversion-like events, and, 
hence, no indication for the existence of this process.
The 90$\%~C.L.$ upper limit for the branching ratio 
$Br(\pi^{0}\rightarrow\gamma + X)$ was calculated by using the following 
relation:

\begin{equation}
N_{X\rightarrow \pi^{0}}^{90\%} > [Br(\pi^{0}\rightarrow\gamma + X)]^2 \times\int \frac{df_{0}(M_{X}, E_{X}, N_{pot})}
{dE_{X}} \sigma_{0}(M_{X},E_{X})  dE_{X} \times  \varepsilon_{sel} \times \rho l \frac{N_{A}}{A}
\end{equation}

where $N_{X-\pi^{0}}^{90\%}$ (= 8.7 events) is the 90$\%~C.L.$ upper limit for
the expected number of signal events,  $\varepsilon_{sel}$ is the selection
efficiency, which was found to be practically independent of $E_X$,
$f_{0}(M_{X}, E_{X}, N_{pot})$ and  $\sigma_{0}(M_{X},E_{X})$ are the flux of
$X$ bosons for the given  number of protons  $N_{pot}$ on the neutrino target
and the cross section for $X \rightarrow \pi^{0}$ conversion on lead,
respectively, calculated for  $Br(\pi^{0}\rightarrow\gamma + X)$=1.\
In Eq.(1), $M_{X}$ and E$_{X}$ are the mass and  energy of the $X$ boson, 
respectively.\ 
The cross section  $\sigma_{0}(M_{X},E_{X})$ is given in  ref.~\cite {11}.\
The total $X -$flux per proton on target calculated as a function of the 
$X$ boson mass  is shown in Figure 5.\
We note that  $Br(\pi^{0}\rightarrow\gamma + X)$ appears twice 
in the formula for $N_{X\rightarrow \pi^{0}}$, through  the $X$ boson flux from
the target and through the  Primakoff mechanism.\ 

The 90 $\% ~C.L.$  branching ratio limit curve is shown in Figure 6
together with the result of ref.~\cite{9}. For the  mass region
 0$<M_{X}<$ 120 MeV/c$^{2}$   the  limit is 
\begin{equation}
Br(\pi^{0}\rightarrow\gamma + X)<(3.3~to~1.9)\times10^{-5} 
\end{equation}
 Varying the cut on the ECAL energy deposition in the range from 5 to 50 GeV 
does not change the limit substantially, while above 50 GeV it becomes worse.
The limit is valid for an $X$ boson lifetime 
$\tau_{X} > 10^{-9} M_{X}[MeV/c^{2}]$ s. For the  mass region 0$<M_{X}<$
60 MeV/c$^{2}$, the limit is approximately a factor 10 to 5 better than the best
previously published limit obtained by the Crystal Barrel 
collaboration~\cite{9}.

Our result can also be used to constrain the magnitude of the coupling of a
hypothetical gauge X boson to quarks~\cite{6,11}
\begin{equation}
 g^{2}< 2\times 10^{-7}(1 - \frac{m^2_X}{m^2_{\pi}})^{-3}
\end{equation} 
 
The attenuation of the $X$- flux due to $X$ interactions with matter was found 
to be negligible, 
since for $Br(\pi^{0} \rightarrow X\gamma) \leq 10^{-4}$ the $X$ boson mean free
path in iron is $\geq$ 300 km, as compared with the iron and earth shielding 
total length of 0.4 km used in our beam. Furthermore the decay length for 
$E_{X} > 10~GeV$ was estimated to be longer than $\simeq$10 km \cite{11}.
We note that limits on the branching ratio 
$Br(\eta, \eta' \rightarrow\gamma + X)$
could also be obtained from our data if the cross-sections for $\eta, \eta'$
productions in the forward direction in $p - Be$ collisions at 450 $GeV$ were
known.

\vspace{.5 cm}

{\large \bf Acknowledgements}

We thank the SPS staff, the neutrino beam group and the technical staffs of the
participating institutions for their vital contributions to the success of this
experiment.

This experiment was supported by
the Australian Research Council and the Department of Industry, Science and Technology, the
Bundesministerium f\"ur Bildung, Wissenschaft, Forschung und Technologie of Germany,
CERN, the 
Institut National de Physique Nucl\'eaire et des Particules and the Commissariat \`a l'Energie Atomique of France, 
the Istituto Nazionale di Fisica Nucleare of Italy,
the Institute for Nuclear Research of the Russian Academy of Sciences and Russian Foundation for Fundamental Research,
the Fonds National Suisse de la Recherche Scientifique,
the U.S. Department of Energy and the U.S. National Science Foundation,
which are gratefully thanked.    
We are indebted to John Ellis and Nikolai Krasnikov for fruitful discussions.

\newpage

\begin{figure}
     \centering
   \mbox{\epsfig{file=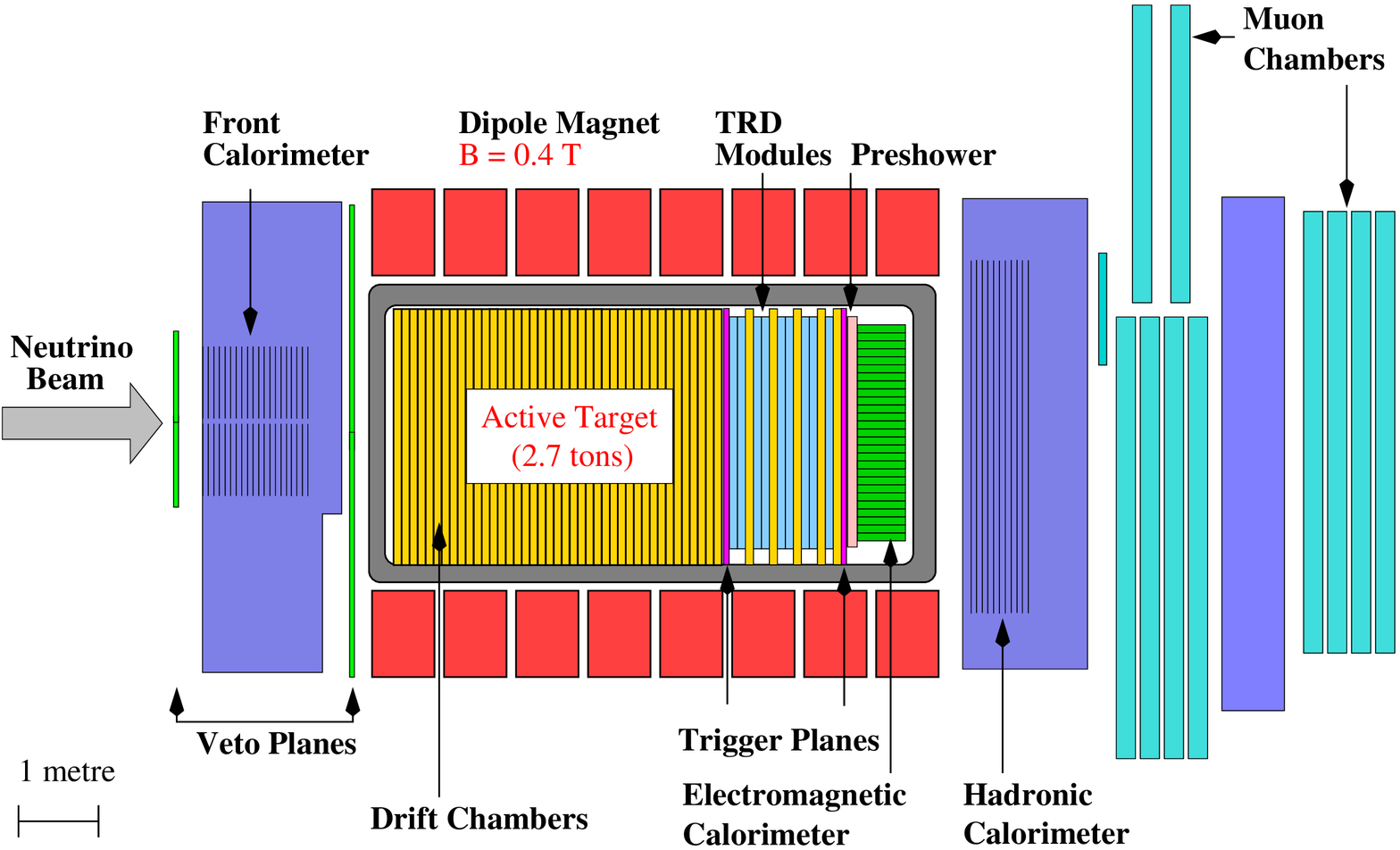,height=170mm,angle=-90}}
     \centering
  \caption{\em Side view of the NOMAD detector.}
  \label{fig:1}
\end{figure}

\newpage
 \begin{figure}
     \centering
   \mbox{\epsfig{file=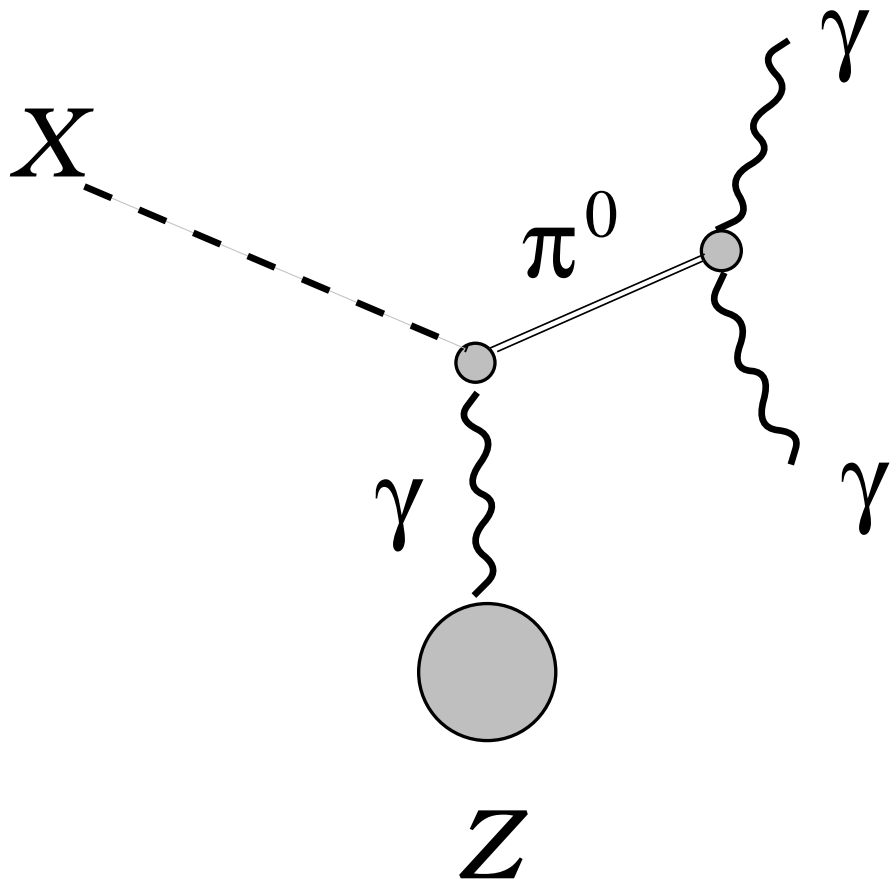,height=120mm}}
    \centering
  \caption{\em Feynman diagram for $\pi^{0}$ production  by Primakoff 
effect.}
  \label{mu1}
\end{figure}

\newpage

 \begin{figure}
     \centering
   \mbox{\hspace{-1.5cm}\epsfig{file=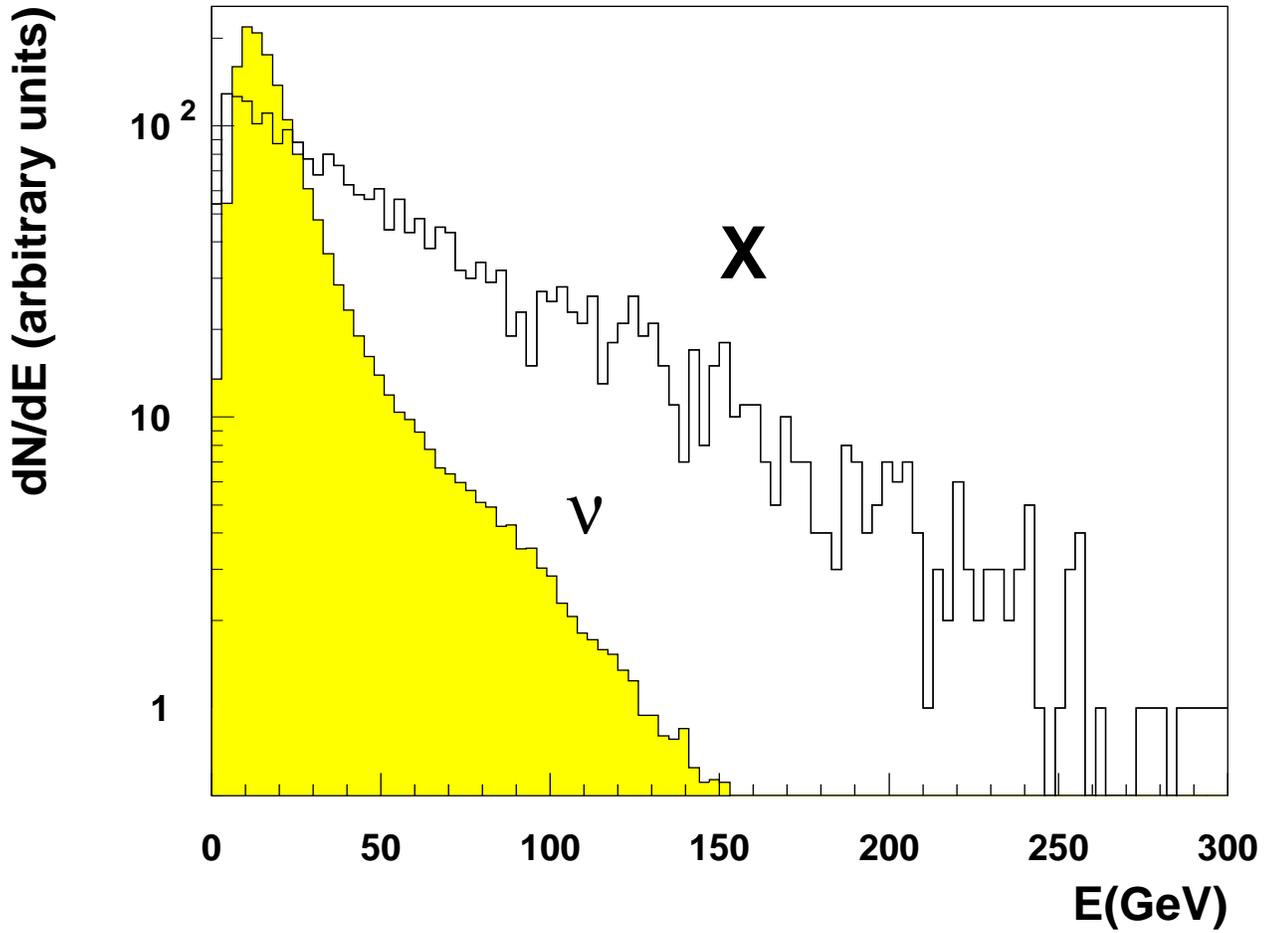,height=150mm}}
    \centering
  \caption{\em Combined energy spectrum of X bosons with mass M$_{X}$=10 MeV from the 
     SPS neutrino target and from the beam dump region at the NOMAD detector.\ The neutrino energy spectrum (hatched area) is also shown for comparison. The
two spectra are arbitrarily normalized.}
  \label{figure 3:}
\end{figure}

\newpage
  \begin{figure}
   \mbox{\hspace{-1.5cm}\epsfig{file=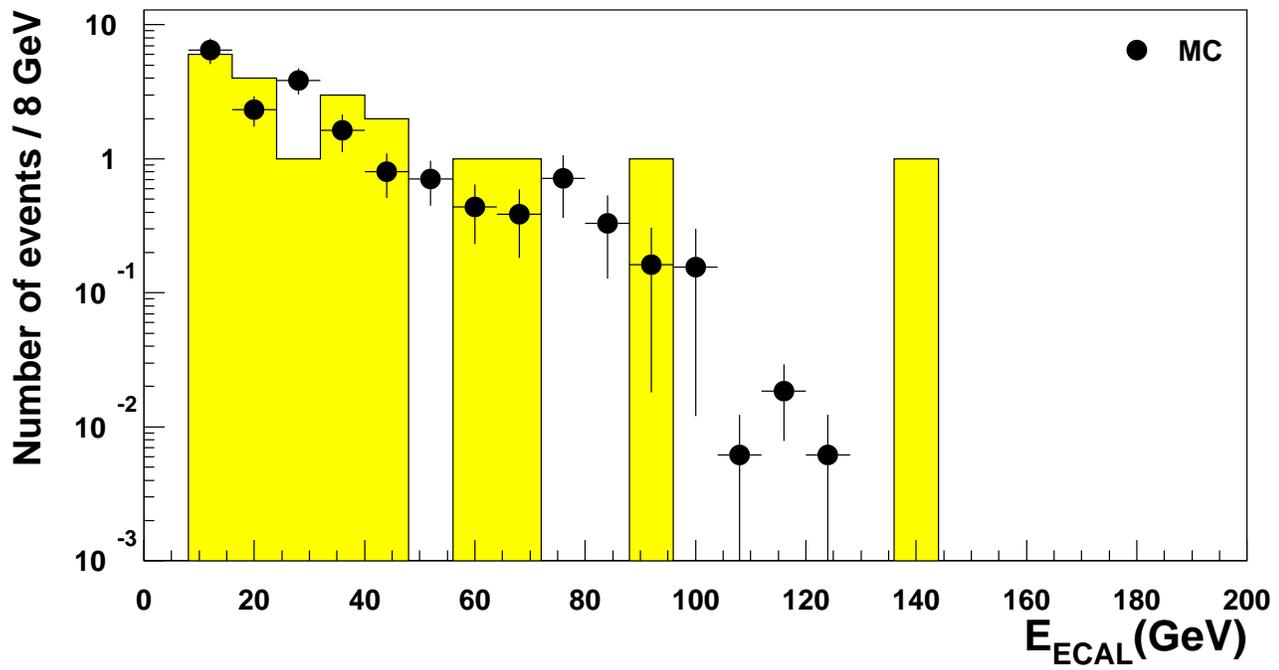,height=110mm}}
    \centering
  \caption{\em ECAL energy distributions for candidate  events (shaded area) 
and for combined background events.}
  \label{figure 4:}
\end{figure}

\newpage

\begin{figure}
  \mbox{\hspace{-1.5cm}\epsfig{file=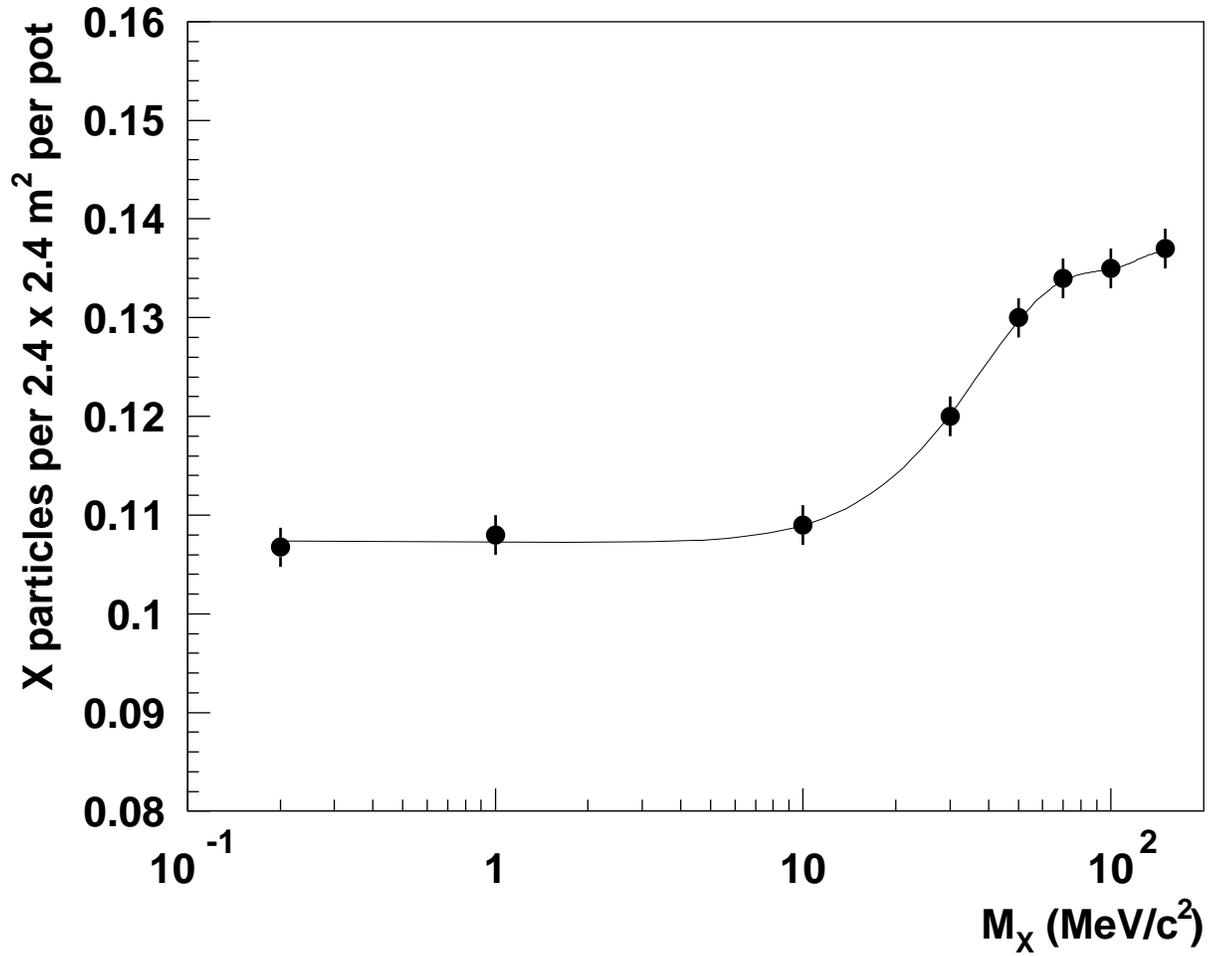,height=150mm}}
  \caption{\em  Total  $X$ boson flux at the NOMAD detector versus mass 
$M_{X}$ ( $E_{X} > 8~ GeV$) calculated for  $Br(\pi^{0}\rightarrow\gamma + X)$=1. The curve is a 
   polynomial fit to the points.}
  \label{figure 5:}
\end{figure}

\newpage
 \begin{figure}
   \mbox{\epsfig{file=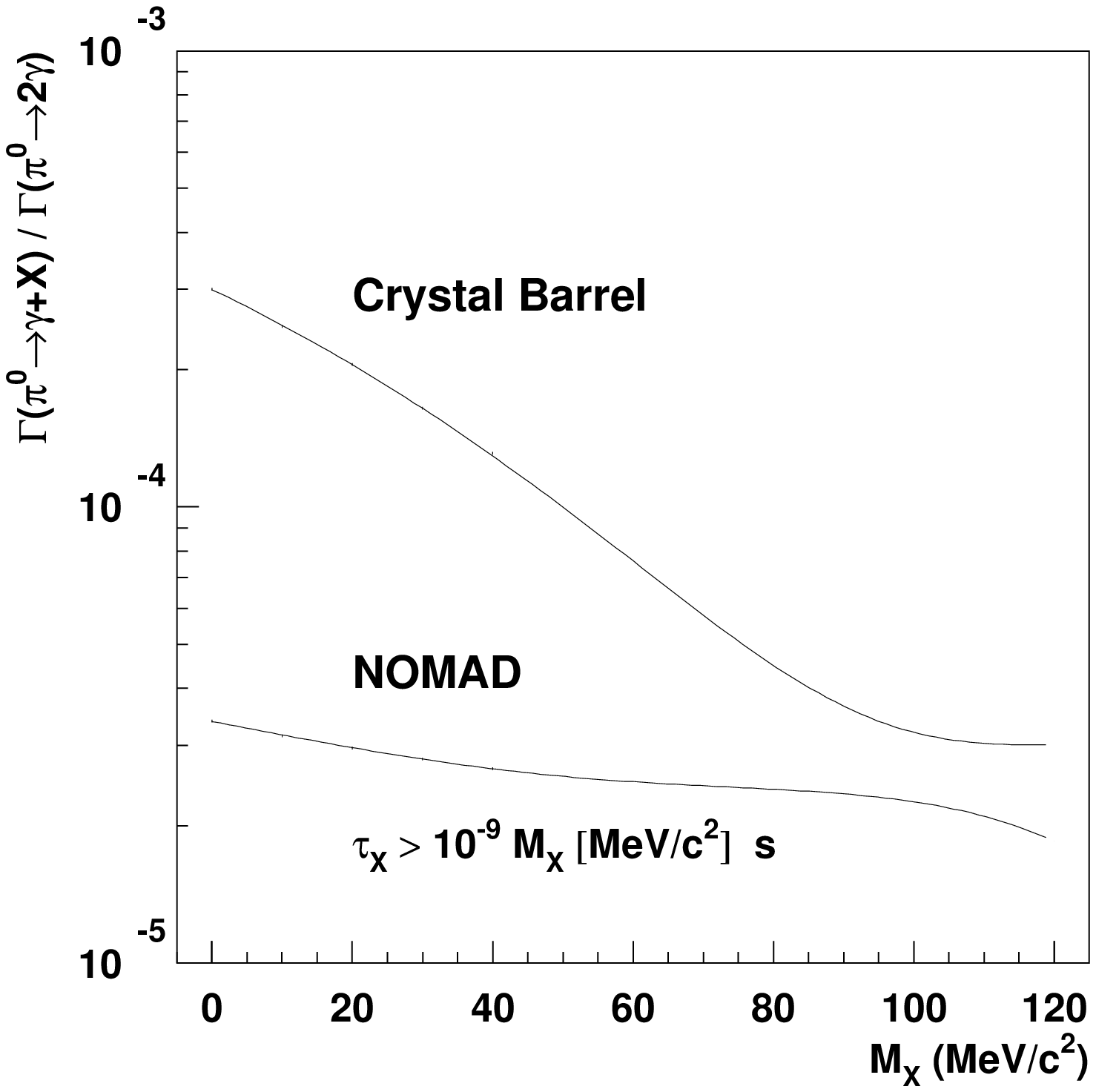,height=170mm}}
    \centering
  \caption{\em  90$\%$ confidence level upper limit on branching 
$Br(\pi^{0}\rightarrow\gamma + X)$ versus $M_{X}$.\ The limit is valid for a lifetime 
 of the $X$ boson $\tau_{X} > 10^{-9} M_{X}[MeV/c^{2}]$
 s.}
  \label{figure 6:}
\end{figure}


\newpage

\begin{thebibliography}{99}
\bibitem{1}
P.~Langacker , Phys.Rep. 
{\bf 72C}(1981)185. 
\bibitem{2}
S.~Weinberg , Phys.Rev.{\bf D26}(1982)287; 

P.~Fayet, Phys.Lett. {\bf 69B}(1977)489, Nucl.Phys.
{\bf B187}(1981)184.
\bibitem{3}
J.~Ellis et al., Nucl.Phys.  {\bf B276}(1986)14.
\bibitem{4}
S.~Glashow, Proc. Conf. Rencontres de Moriond, (1986).
\bibitem{5}
M.I.~Dobroliubov and A.Yu.~Ignatiev, 
Nucl.Phys. {\bf B309}(1988) 655.
\bibitem{6}
M.I.~Dobroliubov, Yad.Fiz. {\bf 52}(1990) 551
[Sov.J.Nucl.Phys. {\bf 52}(1990) 352].

M.I.~Dobroliubov, Z.Phys.  {\bf C 49}(1991) 151.
\bibitem{7} R.~Meijer Drees et al., Phys.Rev.{\bf D49}(1994) 4937.
\bibitem{8} M.S.~Atiya et al., Phys.Rev.Lett.{\bf 69}(1992) 733.
\bibitem{9}
C.~Amsler et al., Phys.Lett.  {\bf B333}(1994) 271;

C.~Amsler et al., Z.Phys.  {\bf C 70}(1996) 219.
\bibitem{10} R.~Meijer Drees et al., Phys.Rev.Lett.{\bf 68}(1992) 3845.
\bibitem{11} S.N.~Gninenko and N.V.~Krasnikov, ''On search for a new light 
gauge boson from $\pi^{0}(\eta)\rightarrow \gamma + X$ decays in neutrino experiments'', hep-ph/9802375, to appear in Phys.Lett.{\bf B}.
\bibitem{12}
J.~Altegoer et al. (NOMAD Collaboration), '' The NOMAD Experiment at the 
CERN SPS'', Nucl.Instr. and Meth. {\bf A404}(1998)96.
\bibitem{13}
M.~Anfreville et al., ''The drift chambers of the NOMAD detector'', to be submitted
to Nucl.Instr. and Meth.(1997).
\bibitem{14} 
G.~Bassompierre et al., Nucl.Instr. and Meth. {\bf A403}(1998)363.\\
G.~Bassompierre et al., ''Performances of the NOMAD Transition Radiation Detector'', LAPP-EXP-97-06, to appear in  Nucl.Instr. and Meth..
\bibitem{15} D.~Autiero et al., Nucl.Instr. and Meth. {\bf A373}(1996)358.
\bibitem{17} GEANT: Detector description and simulation tool, CERN 
Programming Library Long Writeup W5013.
\bibitem{18}
G.~Ingelman, The LUND MC for Deep Inelastic Lepton-Nucleon Scattering,
LEPTO 6.1, Physics at HERA, October 1991;

T.~Sj\"ostrand, Computer Physics Commun. {\bf 39}(1986)347;

H.U.Bengtsson and T.Sj\"ostrand, JETSET, Computer Physics Commun.{\bf 43}(1987)367.
\bibitem{19}  ''Parameterization of $e^{-}$ and $\gamma$ initiated showers
 in the NOMAD lead glass calorimeter'', to be submitted to  Nucl.Instr. and Meth..
\end{thebibliography}
\end{document}